# Pattern recognition with neuromorphic computing using magnetic-field induced dynamics of skyrmions

Tomoyuki Yokouchi[1,2*], Satoshi Sugimoto[3], Bivas Rana[1,4], Shinichiro Seki[1,5,6], Naoki Ogawa[1,5], Yuki Shiomi[2], Shinya Kasai[3,5], Yoshichika Otani[1,7,8]

*Corresponding author. Email: yokouchi@g.ecc.u-tokyo.ac.jp

**Affiliations**

[1]RIKEN Center for Emergent Matter Science (CEMS), Wako 351-0198, Japan

[2]Department of Basic Science, The University of Tokyo, Tokyo 152-8902, Japan

[3]National Institute for Materials Science (NIMS), Tsukuba 305-0047, Japan

[4]Institute of Spintronics and Quantum Information, Faculty of Physics, Adam Mickiewicz University, Poznań, Uniwersytetu Poznanskiego 2, Poznań 61-614, Poland

[5]PRESTO, Japan Science and Technology Agency (JST), Tokyo 102-0075, Japan

[6]Department of Applied Physics and Institute of Engineering Innovation, The University of Tokyo, Tokyo 113-8656, Japan

[7]Institute for Solid State Physics (ISSP), The University of Tokyo, Kashiwa 277-8561, Japan

[8]Trans-scale Quantum Science Institute, University of Tokyo, Bunkyo-ku, Tokyo 113-0033, Japan

**Abstract**

Nonlinear phenomena in physical systems can be used for brain-inspired computing with low energy consumption. Response from the dynamics of a topological spin structure called skyrmion is one of the candidates for such a neuromorphic computing. However, its ability has not been well explored experimentally. Here, we experimentally demonstrate neuromorphic computing using nonlinear response originating from magnetic-field induced dynamics of skyrmions. We designed a simple-structured skyrmion-based neuromorphic device and succeeded in handwritten digit recognition with the accuracy as large as 94.7 % and waveform recognition. Notably, there exists a positive correlation between the recognition accuracy and the number of skyrmions in the devices. The large degree of freedoms of skyrmion systems, such as the position and the size, originate the more complex nonlinear mapping and the larger output dimension, and thus high accuracy. Our results provide a guideline for developing energy-saving and high-performance skyrmion neuromorphic computing devices.

**MAIN TEXT**

**Introduction**

Artificial neural networks, mimicking human brains, exhibit extraordinary abilities in several tasks such as image recognition (*1*), machine translation (*2*), and a board game (*3*). Nowadays, most artificial neural networks rely on silicon-based general-purpose electronic circuits such as a central processing unit (CPU) and a graphics processing unit (GPU). However, these circuits consume a large amount of energy and are approaching the physical limits of downscaling (*4*). Therefore, developing devices specialized for brain-inspired computing, namely neuromorphic devices, is highly required (*4,5*). In particular, nonlinearity and short-term memory effects are essential functions for neuromorphic devices that various spintronic devices can offer (*6,7,8,9,10,11,12,13,14,15,16,17,18,19*). Among them, we focus on a topological spin structure called magnetic skyrmion (*20,21,22,23,24,25,26,27*). So far, skyrmion-based neuromorphic devices such as reservoir computing devices (*9,10,11,12*), synapse devices (*13,14*), and probabilistic computing devices (*15,16*) have been studied to bring about high performance. However, a fully experimental evaluation of its ability for neuromorphic tasks such as pattern recognition is still lacking.

We design the skyrmion neuromorphic computer based on a reservoir computing model (*7,8,9,10,11,28,29,30,31,32,33,34,35*). The conventual reservoir computing model consists of two parts (Fig. 1A). The first part, called the "reservoir part", performs a complex nonlinear transformation of input data into high-dimensional output data. Here, the dimension is the number of linearly independent outputs. In this process, the reservoir part temporally stores the information of past input to make the output depend on both present and past inputs (short-term memory effect). The second part conducts a linear transformation of the outputs from the reservoir part. The coefficient parameters of this linear transformation are optimized by using a training data set so that the final output becomes a desirable one. Incidentally, the nonlinear transformation of input into high-dimensional outputs is the essence of reservoir computing; the linearly inseparable data can become linearly separable in the high-dimensional space, enabling complex data classification as in the kernel method (*36*). Remarkably, optimizing parameters (i.e., training) in reservoir computing is unnecessary for the reservoir part. In other words, the reservoir part performs the complex nonlinear transformation with fixed parameters. Hence, we can implement the reservoir part using a physical system with the complex nonlinearity, memory effect (or equally hysteresis) with short-term properties (*7,8,9,10,11,29,30,31,32,33*). As shown below, skyrmion systems also exhibit nonlinearity and short-term memory effects. Moreover, the skyrmion system has large degrees of freedom because each can take various states with different positions and sizes. This feature theoretically brings about a complex transformation of input data and high performance (*9,10,11*). However, it has not been experimentally explored well. We experimentally found that the skyrmion-based physical reservoir device exhibits good abilities in recognition tasks. Notably, although the structure of the present device is quite simple, the recognition accuracy as high as 94.7 % is obtained in a handwritten digit recognition task, indicating an advantage of the skyrmion system in neuromorphic computing.

**Results**

Our skyrmion-based physical device consists of parallelly connected "subsections" [Fig. 1 (B to D)]. A subsection is a simple-shaped Hall bar made of Pt/Co/Ir film deposited on $SiO_2$/Si substrate, in which skyrmions appear (*37,38,39*). Each subsection has single input and output. The input signal is a time-dependent out-of-plane magnetic field [$H_{AC}(t)$] whose waveform is the same as what we want to compute. The output is anomalous Hall voltage [$V(t)$], which changes in response to $H_{AC}(t)$ because of the $H_{AC}(t)$-induced change in magnetic structures. As shown later, in this process, $V(t)$ depends on the past input signal and is nonlinear to $H_{AC}(t)$ as required in a physical reservoir. Then, we parallelly connect $N$ subsections, in which the different magnitude of a constant out-of-plane magnetic field ($H_{const}$) is applied (Fig. 1D). Because the magnetic structures differ depending on $H_{const}$, the output signals from the subsections tend to be linearly independent of each other. We input the same signal into $N$ subsections. Hence, the skyrmion-based neuromorphic computer device nonlinearly converts a one-dimensional time-series input [$H_{\text{AC}}(t)$] to a linearly-independent $N$-dimensional time-series outputs [$\boldsymbol{V}(t) \in \mathbb{R}^N$] as follows:

$$H_{\text{AC}}(t) \to \boldsymbol{V}(t)$$

$$\boldsymbol{V}(t) = [V^1(t), \cdots, V^N(t)].$$

Here, $V^i(t)$ is the output signal of the $i$-th subsection. This nonlinear mapping into the high-dimensional space is crucial for skyrmion-based neuromorphic computing like conventional reservoir computing. The final output is a linear combination of sampling data from $\boldsymbol{V}(t)$. The linear combination coefficients are optimized using a training data set to ensure the final output is desirable (see Method for details). We used only one Hall bar in the actual measurement and obtained $V(t)$ by repeating the measurement $N$ times for the same $H_{\text{AC}}(t)$ with $H_{\text{const}}$ changed. All experiments were performed at room temperature.

First, we present basic properties of response in a single subsection, which shows short-term memory effect and nonlinearity. Figure 2 (A to D) shows the time dependence of the input magnetic field $H_{\text{AC}}(t)$ and the output anomalous Hall voltage $V(t)$. When we applied two cycles of a sine-wave magnetic field, $V(t)$ exhibited distinct variation [Fig. 2C]. This change originates from the magnetic field-induced transformation of the spin structure. As shown in Fig 2 (E to H), the size, form, and the number of skyrmions vary in response to $H_{\text{AC}}(t)$, and consequently, the total magnetization in the Hall bar area also varies, which leads to the observed change in $V(t)$. Besides, the $V(t)$ signal depends on a past input; when we change the first cycle of the input signal from the sine wave to a square wave, as shown in Fig.2 B, the time profile of $V(t)$ differs from that in the case of two cycles of the sine wave [see Fig. 2 (C and D)]. In particular, the second cycle of the input signal is a sine wave in both cases; however, $V(t)$ profiles corresponding to the second cycle are significantly different. In other words, the output signal depends on the past input signal (i.e., memory effect). This memory effect is due to the history-dependent time evolution of the spin structures originating from the first-order transition nature of the skyrmion system, namely topological protection. As shown in Fig. 2 (E to L), the time evolution of spin structure for the first cycle differs between the sine and square waves [Fig. 2 (F and J)]. As a result, the spin structure during the second cycle of $H_{AC}(t)$ is also different in two cases [Fig. 2 (G and K)], which makes $V(t)$ depend on past inputs.

Moreover, the memory effect in the skyrmion system has a short-term property. In other words, after turning off the input signal, the output signal fades out and goes back to an initial value. As shown in Fig. S1, after two cycles of the sine wave are input, the output signals start to return to an intimal value, which indicates that the skyrmion system has the short-term memory property. We note that the time to return the initial state depends strongly on $H_{const}$, and in some $H_{const}$ values, the time is more than several tens of seconds. We also investigate the nonlinearity in $V(t)$ vs. $H_{AC}(t)$. We measured the output anomalous Hall voltage $V(t)$ when two cycles of sin wave magnetic field with various amplitudes were applied [Fig. 2 (M and N)]. The magnitude of $V(t)$ at $t = 2.5$ s as a function of the amplitude of the input magnetic field is presented in Fig. 2 O. The $V(t)$ is not proportional to the amplitude and even shows the sign change, which indicates the strong nonlinearity of the output anomalous Hall voltage to the input magnetic field.

Next, we demonstrate a waveform classification task, widely used as a benchmark task for neuromorphic computing (*7,34,35*). In this task, the input signal is a waveform of a random combination of sine and square waves (Fig. 3A), and the desired output is 1 for the sine waves and −1 for the square waves. We input the waveform in Fig. 3A into the skyrmion-based reservoir device with $N = 41$ subsections. The amplitude of $H_{AC}(t)$ is 24 Oe. Before inputting the signal, we create the ferromagnetic state by applying a large magnetic field to erase the memory of previous inputs. As an example, $V^i(t)$ signals outputted from some subsections ($H_{const}$ = 1.04, 0.00, and -1.60 Oe) are displayed in Fig. 3B; the input data is complicatedly transformed, and their profiles differ in different subsections as expected. We sampled the output signals with the sampling rate of 100 Hz and calculated the final output as $y(t_k) = \sum_{i=1}^{41} W_i V^i(t_k)$, where $W_i$ ($i = 1, 2, \cdots, 41$) is time-independent coefficients, and $t_k$ is the time at the $k$-th sampling point (see also Method). Then, $W_i$ is optimized by using the first half of the waveform (0 to 25 s) so that the mean squared error between $y(t_k)$ and the target value is the minimum, and finally, we binarized $y(t_k)$ (see Method for details). As shown in Fig. 3 (C and D), the output values follow well the desired output values in not only the input data set used for the training (0-25 s) but also not used for the training (25-50 s).

To investigate the effect of the skyrmion formation on the recognition accuracy, we fabricate the Hall-bar devices accommodating fewer skyrmions and ferromagnetic-like domains (Device B to D) as shown in Fig. 3E by controlling the strength of Dzyaloshinsky-Moriya interaction (DMI) and perpendicular magnetic anisotropy (see Method). Here, the magnitude of DMI gradually decreases from device A to device D. Then, we performed the same waveform recognition task with various amplitudes of $H_{AC}(t)$. As shown in Fig. 3F, the recognition accuracy in device A, which has the largest skyrmion population of the four devices, is high. In contrast, device D, which accommodates ferromagnetic domains, exhibits low recognition accuracy for all $H_{AC}(t)$ amplitudes. We count the number of skyrmions existing during the waveform recognition task $<n_{sk}>$ (see Method for details); as shown in Fig. 3G, we found a large $<n_{sk}>$ in device A and a small number in device D as expected. Figure 3H shows the correlation between the recognition accuracy and $<n_{sk}>$, exhibiting a positive correlation. This result suggests that the skyrmion formation is critical in improving recognition accuracy.

Before discussing the origin of better recognition accuracy in the skyrmion system, we demonstrate that the skyrmion-based reservoir device can solve a more complex task: handwritten digit recognition. We employ the commonly used Mixed National Institute of Standards and Technology database (MNIST) (*40*), some examples of which are shown in Fig. 4F. A preprocessing was performed to convert a two-dimensional image to a one-dimensional input signal (see Method for details). Figure 4 (A to D) shows the preprocessing for an input digit "5", as an example. We input the converted signal into the skyrmion-based reservoir device (device A) with $N = 9$ subsections. The output signals corresponding to the input digit "5" are presented in Fig. 4E. The final output is obtained by a linear transformation of the output signals from each subsections (see Method for details). Using 13219 train images, we optimize the coefficients of the linear transformation. After the optimization, 5000 test images not included in the train data set are used to test the recognition accuracy. Figure 4G presents a confusion matrix obtained in the test process, which shows that the skyrmion-based reservoir well outputs desired digits. The recognition accuracy is $94.7 \pm 0.3$ %. This accuracy is better than an experiment in WO$x$ memristors-based reservoir device (88.1 %) (*29*), simulation in a nanowire-based reservoir system (90 %) (*33*), and chip-level simulation in a skyrmion-based artificial synapse system (89 %) (*13*).

**Discussion**

Finally, we discuss the origin of the better recognition accuracy obtained using the skyrmion-based neuromorphic device than the ferromagnetic domain-based one. First, the creep motion of ferromagnetic domains decreases the recognition accuracy. In Fig. S2, we present the 41 output signals ($H_{\text{const}}$ = -1.6 Oe to 1.6 Oe) in the waveform recognition task for skyrmions (device A) and ferromagnetic domains (device C). In the case of ferromagnetic domains, the center of the oscillation (the red lines in Fig. S2 D) gradually changes with time at low $H_{\text{const}}$. This tendency originates from a slow change in the total magnetization in the Hall bar due to the thermally induced creep motion of the ferromagnetic domains. Such a gradual change in the background must reduce the recognition accuracy because even if we input the same signal, the outputs might be different depending on time, causing false recognition. In contrast, in the case of skyrmion (device A), the output signals oscillate around the time-independent values. This is because thermal agitation has a low impact on the number of skyrmions due to the topological stability of skyrmions. In other words, the topological stability surpasses the thermal agitation; the profiles of the output signals are determined by the form of the input signal and reproducible, which contributes to a high recognition accuracy in the skyrmion devices.

Second, the more significant number of output data dimensions, which originates from the large degree of freedoms of the skyrmion system, also contributes to better recognition accuracy. As mentioned above, the complex nonlinear mapping into high-dimensional space is a crucial factor for the present neuromorphic computing. Because of the particle nature of skyrmions, skyrmion systems have many degrees of freedom, such as position and skyrmion size, causing different spin structural responses to the input signals $H_{\text{AC}}(t)$. This results in high-dimensional mapping. However, the ferromagnetic domain state consists of only two internal states (up and down domains). Hence, the transformation should be less complex than the skyrmion system. To further discuss the dimensionality, we evaluate the dimensionality of the experimentally obtained output signals. The dimensionality is defined by the linearly independent outputs from the subsections. Thus, we plot an output signal of the *i*-th subsection ($V^i$) obtained in the waveform recognition task as a function of an output signal of the *j*-th subsection with a different $H_{\text{const}}$ value ($V^j$) ($i \neq j$). If $V^i$ and $V^j$ are linearly dependent (i.e., $V^i=CV^j$, where *C* is a coefficient), the profile becomes a straight line. However, if $V^i$ and $V^j$ are linearly independent, the curve shape becomes non-monotonous. As shown in Fig. S3 A, the profiles for the skyrmion-based device tend to be non-monotonous smooth curves. In contrast, the ferromagnetic domain-based device profiles are relatively straight and squarish (Fig. S3 B). These results indicate that the number of linearly independent outputs in the skyrmion-based device is more than that in the ferromagnetic domain device. In other words, the skyrmion-based device has a larger dimensionality than the ferromagnetic domain-based device. This fact contributes to the better recognition accuracy in the skyrmion-based device.

We experimentally conclude that the skyrmion system is a promising candidate for neuromorphic computing. The high degree of freedom and topological stability of the skyrmions lead to reproducible, complex, and high-dimensional mapping and, consequently, better recognition accuracy. The present skyrmion-based neuromorphic system consists of less than ten simple-shaped and microscale Hall bars. Nevertheless, the recognition accuracy in the handwritten digit recognition task is better than other neuromorphic devices (*13,29,33*), which require the fabrication of a large number of nanoscale objects. Our findings provide a novel pathway for designing a high-performance neuromorphic computer.

**Materials and Methods**

**Device fabrications**

We deposited multilayer films on $SiO_2$/Si substrates by direct current (DC) and radio-frequency (RF) magnetron sputtering. The complete stack structure of the films used in this work is $SiO_2$/Si substrates/Ta(5 nm)/Pt(5 nm)/Co($d_{Co}$)/Ir(0.8 nm)/Pt(5 nm) in which the nominal thickness of Co ($d_{Co}$) is gradually varied from 0.5 nm to 0.6 nm. The Co layer was deposited by using DC sputtering, and the other materials were deposited by RF sputtering. Thermodynamically stable skyrmions form in an area with $d_{Co}$ ~ 0.55 nm and ferromagnetic domains are observed in a thicker area. This is because $d_{Co}$ affects the perpendicular magnetic anisotropy and the magnitude of DMI, which determine the stable spin structure (*37,38,39*). Next, we patterned the films by using maskless UV lithography followed by Ar ion milling. The width of the Hall bars is 40 μm

**Waveform recognition**

Waveform recognition task is divided into two parts: (1) the transformation of an input signal by using the physical neuromorphic device and (2) a linear transformation of the signals output from the neuromorphic device and optimization of their weights.

For the first part, we input a time-dependent magnetic field $H_{const}$+$H_{AC}$(*t*). Here, $H_{const}$ is a constant out-of-plane magnetic field in order to make a magnetic structure in each subsection different one, which results in the linearly independent outputs as mentioned in the main text. The wave profile of $H_{AC}(t)$ corresponds to the signal which we want to process and is a random combination of sine and square waves as shown in Fig. 3A in this case. We generate $H_{const}$+$H_{AC}$(*t*) by applying the current to a coil with the use of a function generator (NF WF1974) and a bipolar amplifired (NF HSA42011). The direction of magnetic field is perpendicular to the film plane. The consequent time-dependent Hall voltage [*V*(*t*)] is measured with lock-in techniques (NF LI5650) by applying AC current with current density $j = 2.5 \times 10^8$ $Am^{-2}$ and frequency *f* = 333 Hz. The skyrmion-based neuromorphic computer used here has 41 subsections with different $H_{const}$ values. Thus, we obtain the converted signals as follow: $\boldsymbol{V}(t_k) = [V^1(t_k), \cdots, V^{41}(t_k)]$. Here,

$V^i(t)$ is the output signal of the $i$-th subsection and $t_k$ is time at the $k$-th sampling point. We sampled the data with the frequency of 100 Hz (i.e. $k = 1, 2, \cdots, 2000$ and $t_0 = 0, t_1 = 0.01, \cdots, t_k = 0.01k, \cdots, t_{5000} = 50$ s ).

The second part are performed on a conventional computer. The final output signal is the linear combination of $V(t_k)$ as follows $y(t_k) = \sum_{i=1}^{41} W_i V^i(t_k)$. Here, $W_i$ is time-independent coefficient and is optimized to minimize the mean square error $\sum_{k=1}^{2500}[y(t_k) - L(t_k)]^2$. The label $L(t_k)$ is 1 when the input signal is the sine wave and -1 the square wave [Fig. 3 (C and D)]. For the optimization, we use the first half of the data set (i.e. 2500 data obtained from $t = 0$ to 25 s). Finally, we binarized the $y(t_k)$.

**Count of the skyrmion number during the waveform recognition task**

During the waveform recognition task, we take a video of time-evolution of magnetic contrast with the use of a poler Kerr microscopy. The exposure time is approximately 50 ms. Then, we count the number of skyrmions by using a conventional binarization method. Here we defined a particle whose size is below $6 \times 6$ μm$^2$ as skyrmion. Finally, the number of skyrmions is normalized by the area of the Hall bar and the execution time of the task, and <$n_{sk}$> is obtained.

**Handwritten digit recognition**

The handwritten digit recognition task is divided into three parts: (1) transformation of two-dimensional data into one-dimensional input data (2) nonlinear transformation of an input signal by using the skyrmion-based reservoir computing device, and (3) a linear transformation of the output signals from the reservoir and optimization of their weights.

At the beginning of the first part, we removed the unused border area of the images by reducing the original 28×28 images into a 22×20 image (Fig. 4A). Then, we reshaped the two-dimensional 22×20 data points to one-dimensional 440 data points (Fig. 4B). In the next step, we multiply each data point by one cycle of sin wave consisting of 20 data points and obtained a sequence of 440 modulated sin waves [Fig. 4(C and D)]. (i.e. the number of total data points is 8800).

In the second part, we input a time-dependent magnetic field $H_{\text{const}} + H_{\text{AC}}(t)$ into the skyrmion-based neuromorphic device with 9 different $H_{\text{const}}$ values. The amplitude of $H_{\text{AC}}(t)$ is 64 Oe for 4 subsections and 70 Oe for 5 subsections. The frequency is 200 Hz. The consequent time-dependent Hall voltage [$V(t)$] is measured with lock-in techniques by applying AC current with current density $j = 2.5 \times 10^8$ Am$^{-2}$ and frequency $f = 2.99$ kHz. The sampling rate is 80 Hz. The output signal for $d$-th data then is obtained as follow: $\boldsymbol{V}_d = \left(V_d^1(t_k), \cdots, V_d^i(t_k), \cdots, V_d^9(t_k)\right)$, where $t_k$ is time at the $k$-th sampling point, i.e. $k = 1, 2, \cdots, 176$ and $t_0 = 0, t_1 = 0.0025, \cdots, t_k = 0.01k, \cdots, t_{172} = 2.2$ s ).

Finally, we calculate the final output for $d$-th data set as following:

$$\boldsymbol{y}^d = \boldsymbol{V}_d W$$

$$= \left(V_d^1(t_0), \cdots V_d^1(t_{176}), \cdots, V_d^9(t_0), \cdots, V_d^9(t_{176})\right) \begin{pmatrix} W_{1,1} & \cdots & W_{1,10} \\ \vdots & \ddots & \vdots \\ W_{1584,1} & \cdots & W_{1584,10} \end{pmatrix}.$$

Here, the output $\boldsymbol{y}^d$ is $1 \times 10$ vector and W is $1584 \times 10$ weight matrix. Then, we introduce $1 \times 10$ label vector $\boldsymbol{L}^d$. If the $d$-th input data is a digit of $m$, the $m$-th components of $\boldsymbol{L}^d$ is one and the others zero. We optimize to minimize the mean square error $\sum_{d=1}^{13219}\|\boldsymbol{y}^d - \boldsymbol{L}^d\|^2$ by using 13219 train data. In the test process, we calculate the $\boldsymbol{y}^d$, and the predicted digit is determined from the maximum component of $\boldsymbol{y}^d$. For evaluation of the recognition accuracy, 5000 test images that are not included in the train data set are used. In order to reduce the ambiguity due to a choice of test and train data, we repeated optimization 10 times with different choices of test and train data and take the average.

**Acknowledgments**

**Funding:** This work was supported by the JSPS Grants-in-Aid for Scientific Research (A) (Grant Nos. 18H03685, 20H00349, 21H04440), Scientific Research (B) (Grant No. 21H01794), and Young Scientists (Grant No. 19K14667), by JST PRESTO (Grant Nos JPMJPR18L3, JPMJPR18L5, and JPMJPR17I3) by Asahi Glass Foundation and by Murata Science Foundation.

**Author contributions:** YO and TY conceived the project. S.Sugimoto and SK deposited the thin films. TY fabricated the devices, conducted the measurements with assistance from B.R., S.Y., S.Seki, and NO. TY analyzed the data and performed the calculation. TY and YO wrote the draft. All authors discussed the results and commented on the manuscript.

**Competing interests:** The authors declare no competing financial interests.

**Figures and Tables**

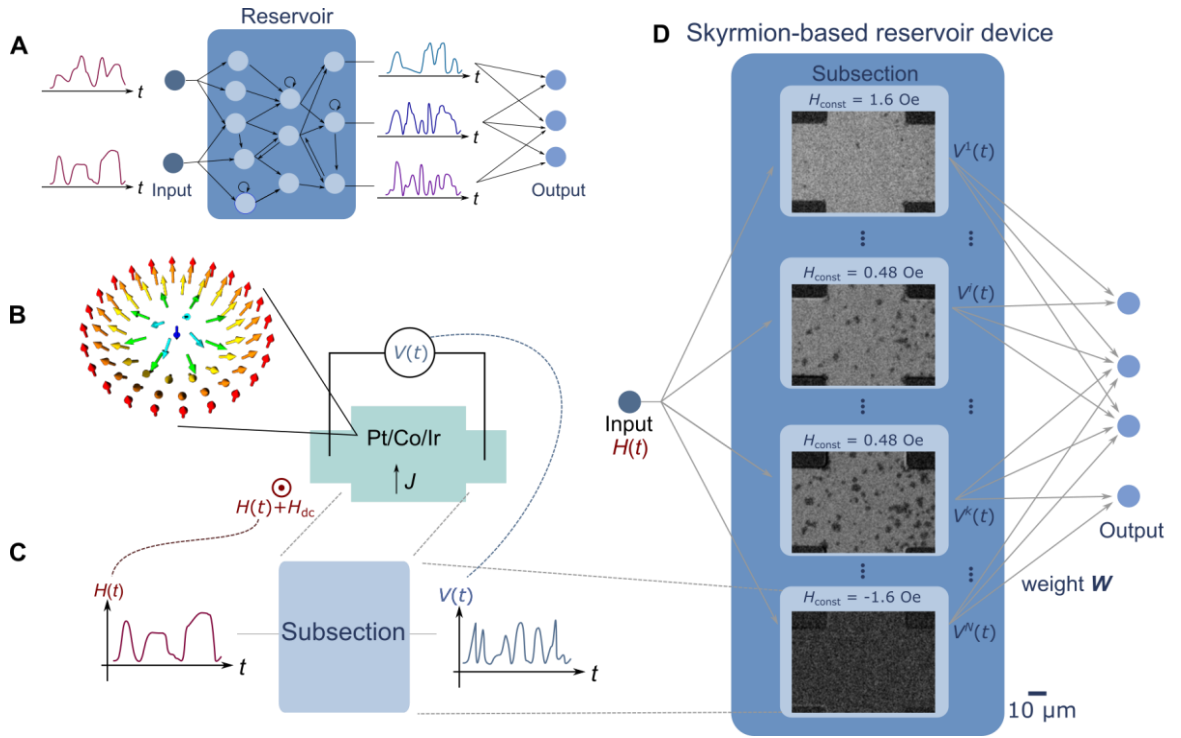

**Fig. 1. Concept of skyrmion-based neuromorphic computing.** (**A**) Schematic for conventional reservoir computing model. (**B**) Schematic illustration of a Hall bar device and a magnetic skyrmion. (**C**) Conceptual diagram for the data conversion in a subsection. (**D**) Schematic illustration of a skyrmion-based neuromorphic computer. Polar Kerr images of subsection with various constant magnetic field ($H_{\text{const}}$) in the absence of time-dependent magnetic field [$H_{\text{AC}}(t)$] are also presented.

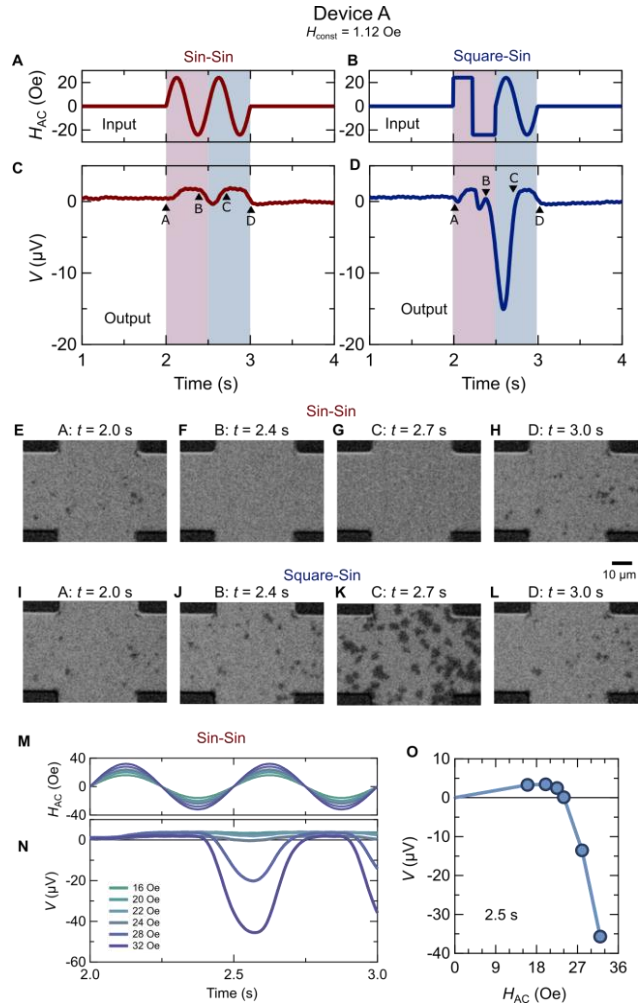

**Fig. 2. Memory effect and nonlinearity in the skyrmion system.** (**A** to **D**) The time profile of input magnetic field ($H_{AC}$) (A, B) and corresponding Hall voltage (C, D) in the Hall-bar device A with the constant magnetic field $H_{const}$ = 1.12 Oe. (**E** to **L**) Snapshots of polar Kerr images during applying $H_{AC}$ for the Sin-Sin input (E to H) and the Square-Sin input (I to L). The corresponding time points are represented in (C, D) by the triangles. (**M** and **N**) The time profile of $H_{AC}$ with various amplitudes (M) and corresponding Hall voltage output (N) in the Hall-bar device A with the constant magnetic field $H_{const}$ = 1.12 Oe. (**O**) The Hall voltage output at $t$ = 2.5 s as a function of the amplitude of $H_{AC}$. The solid line is a guide for eyes.

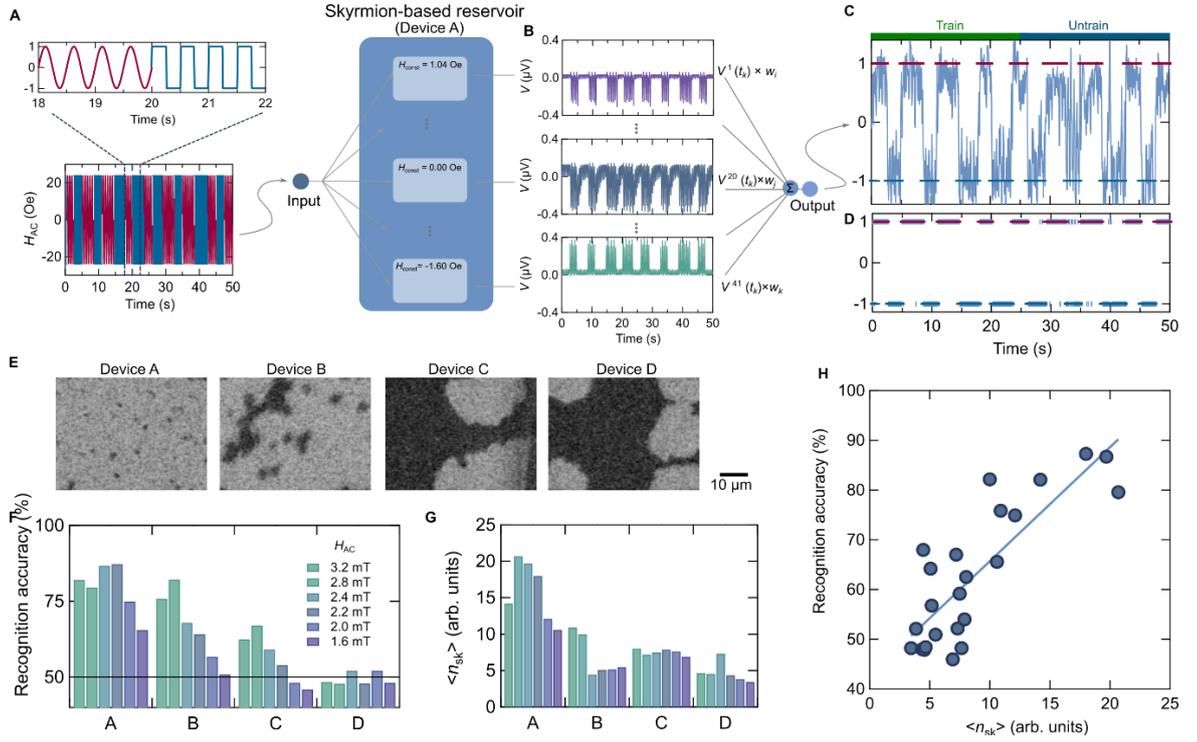

**Fig. 3. Waveform recognition task.** (**A**) The waveform of input signal [$H_{AC}(t)$] for waveform recognition task. The input signal is a waveform of a random combination of sine (blue) and square waves (red). (**B**) The output signals (*V*) in some of the subsections with different constant magnetic fields ($H_{const}$). (**C** and **D**) The final output calculated by the linear combination of the output signals in the 41 subsections (C) and its binarization (D). The blue and red lines are desired output. (**E**) Polar Kerr images of device A to D. (**F**) The recognition accuracy in the waveform recognition task for device A to D for various amplitudes of $H_{AC}$. (**G**) The average number of skyrmions <$n_{sk}$> during the waveform recognition task for device A to D for various amplitudes of $H_{AC}$. (**H**) A scatter plot of <$n_{sk}$> and the recognition rate. A correlation coefficient of 0.82 is obtained.

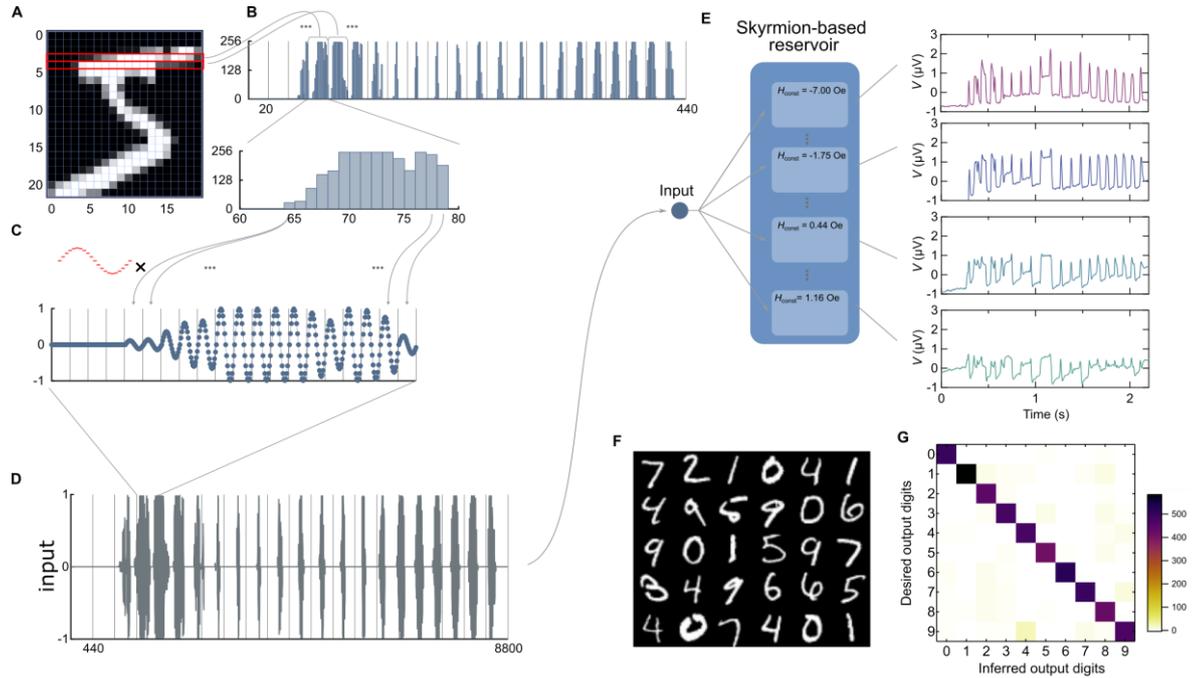

**Fig. 4. Handwritten digits recognition task.** (**A-D**) Schematic for preprocessing for handwritten digits recognition task. The two-dimensional image (A) is converted to a one-dimensional array (B). A sin wave is multiplied by each data point (C), and an input signal is obtained (D). (see also Method for details) (**E**) The output signals from subsections corresponding to the input digit "5". (**F**) Some examples from the MNIST database. (**G**) A confusion matrix showing the recognition results from the skyrmion-based reservoir vs the desired outputs. Recognition accuracy of 94.7±0.3 % is obtained.

# Supplementary Materials for

# Pattern recognition with neuromorphic computing using magnetic-field induced dynamics of skyrmions

Tomoyuki Yokouchi[*], Satoshi Sugimoto, Bivas Rana, Shinichiro Seki, Naoki Ogawa, Yuki Shiomi, Shinya Kasai, Yoshichika Otani

*Corresponding author. Email: yokouchi@g.ecc.u-tokyo.ac.jp

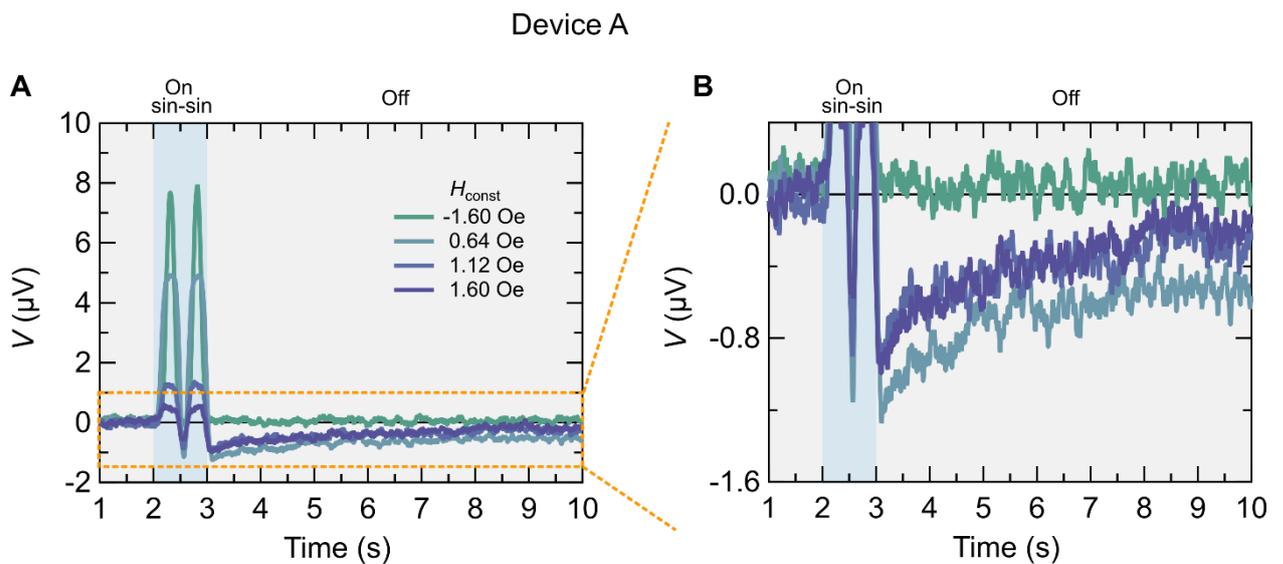

**Fig. S1. Short-term memory effect in the skyrmion system** (**A**) The Hall voltage as a function of time when the two cycles of sin wave with the amplitude of 24 Oe is applied between 2.0 to 3.0 sec. (**B**) A magnified view of the region marked by the orange square in (A).

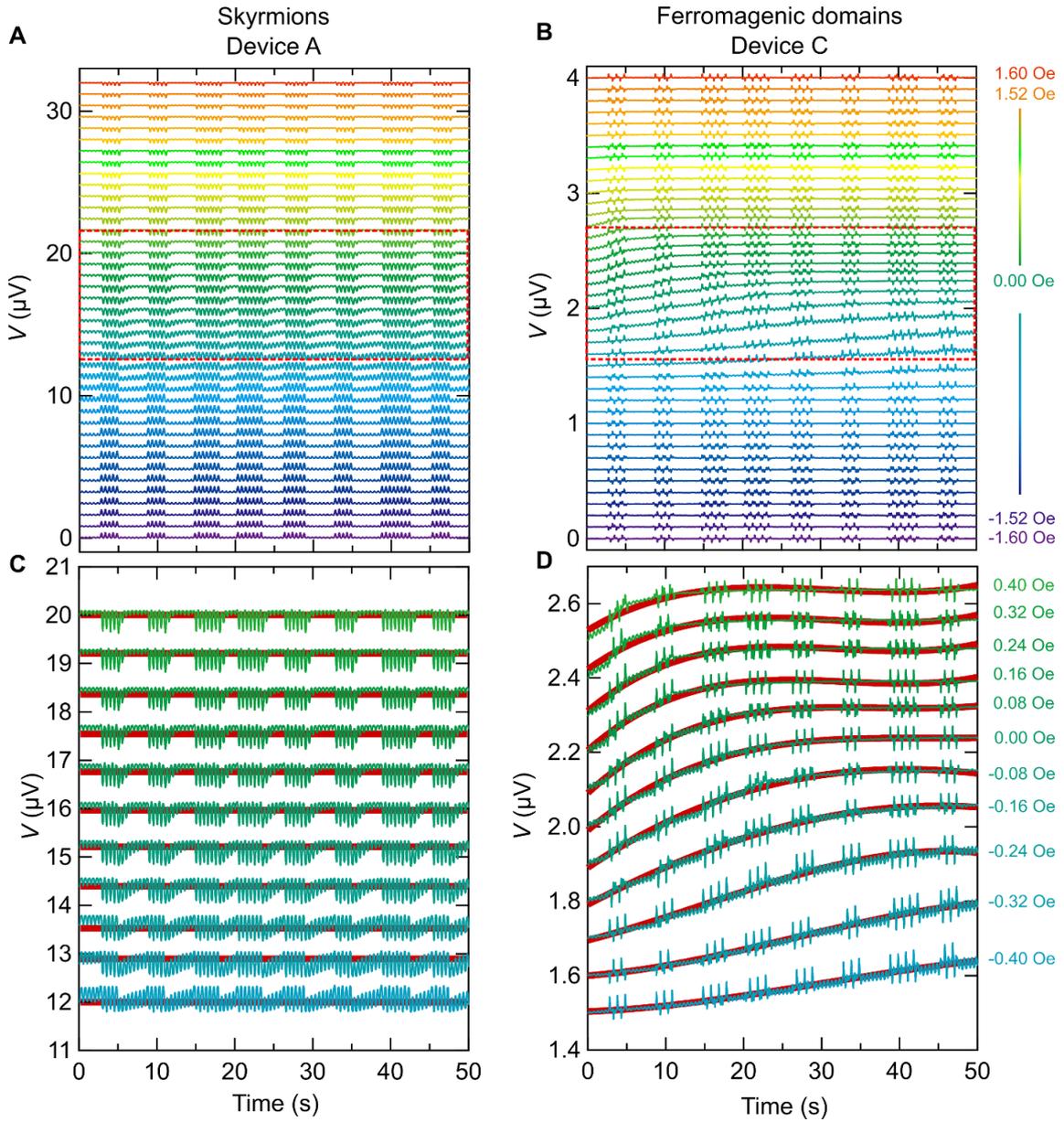

**Fig. S2. Output signals for the waveform recognition task.** (**A** and **B**) The output signals for the waveform recognition task of all the subsections in the skyrmion-based device (A) and the ferromagnetic domain-based device (B). The amplitude of $H_{AC}(t)$ is 24 Oe. (**C** and **D**) A magnified view of the region marked by the red square in (A and B). The red lines are guides for eyes.

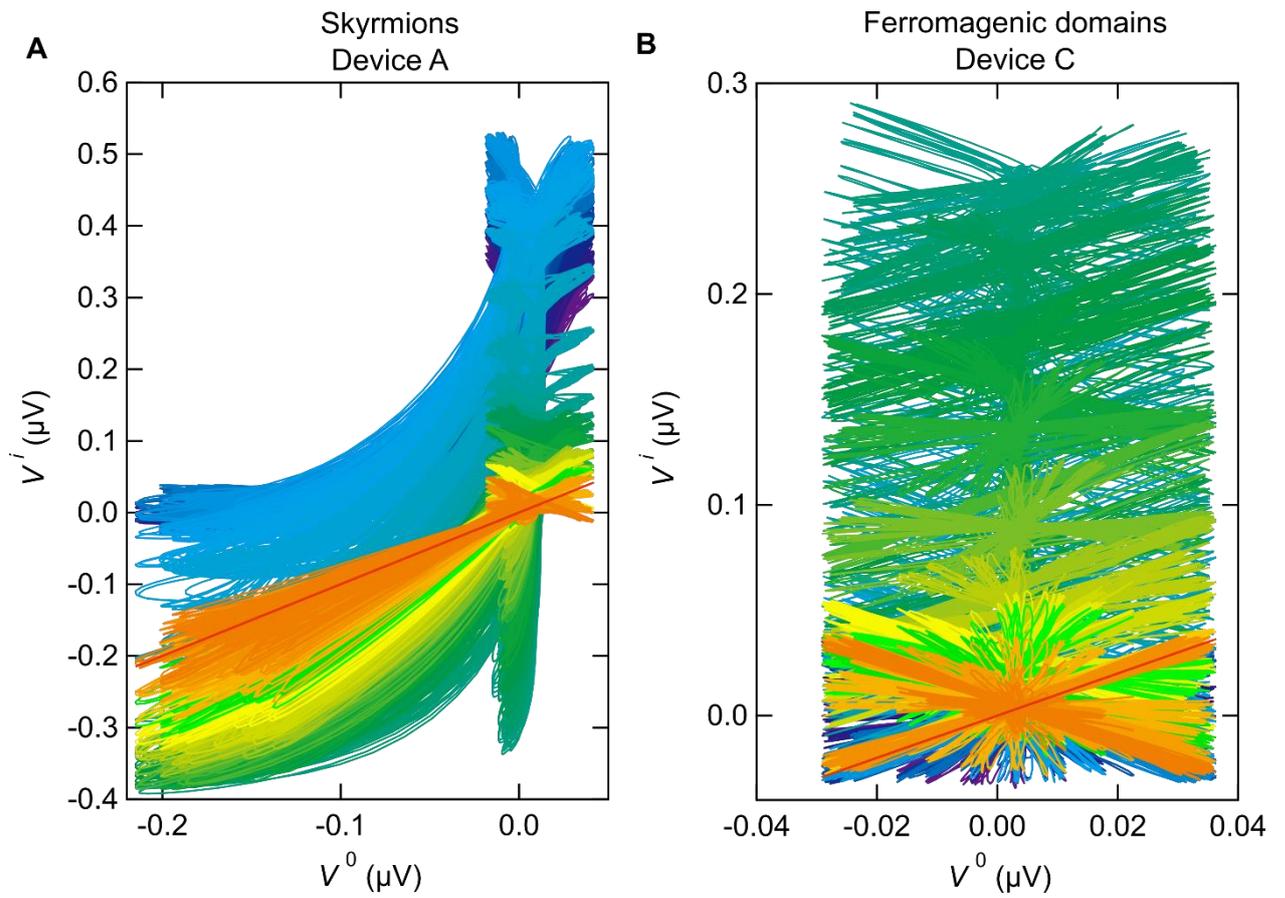

**Fig. S3. Dimensionality of the output signals.** (**A** and **B**) The output signals of $i$-th subsection ($V^i$) with various $H_{const}$ plotted as a function of the output signal of the subsection with $H_{const}$ = 1.6 Oe ($V^0$) in the skyrmion-based device (A) and the ferromagnetic domain-based device (B).